\documentclass[preprint,12pt]{article}
\usepackage{amssymb}
\usepackage{amsmath}
\usepackage{hyperref}

\begin{document}

\centerline{}

\centerline {\Large{\bf The nonlocal Darboux transformation of
 }}

\centerline{}

\centerline{\Large{\bf the 2D stationary Schr\"odinger equation
and }}

\centerline{}

\centerline{\Large{\bf its relation to the Moutard
transformation}}

\centerline{}

\centerline{\bf {A. G. Kudryavtsev}}

\centerline{}

\centerline{Institute of Applied Mechanics,}

\centerline{Russian Academy of Sciences, Moscow 119991, Russia}

\centerline{kudryavtsev\_a\_g@mail.ru}

\begin{abstract}

The nonlocal Darboux transformation of the two - dimensional
stationary Schr\"odinger equation is considered and its relation
to the Moutard transformation is established. It is shown that a
special case of the nonlocal Darboux transformation provides the
Moutard transformation.  New examples of solvable two -
dimensional stationary Schr\"odinger operators with smooth
potentials are obtained as an application of the nonlocal Darboux
transformation.

\end{abstract}

\centerline{PACS: 02.30.Jr, 02.30.Ik, 03.65.Ge}






\section{Introduction}

The two - dimensional stationary Schr\"odinger equation is a model of physical
 phenomenon in nonrelativistic quantum mechanics \cite
{Landau}, acoustics \cite {Morse} and tomography \cite {Kak}.
The case of fixed energy for two - dimensional Schr\"odinger equation is of interest
 for the theory of two - dimensional integrable nonlinear systems \cite
{Veselov1984}, \cite {Novikov2010} and for the multidimentional inverse scattering theory \cite {Grinevich2000} .

Consider the two - dimensional stationary Schr\"odinger equation in the form
\begin{equation} \label{eq1}
\left( \Delta -{\it u}\left( x,y \right) \right) Y \left( x,y
\right) =0
\end{equation}

where $\Delta$ is the two - dimensional Laplace operator
\begin{equation*}
\Delta= {\frac {\partial ^{2}}{\partial {x}^{2}}}
+{\frac {\partial ^{2}}{\partial {y}^{2}}}
\end{equation*}

The Schr\"odinger equation is related with the Fokker-Planck equation

\begin{equation} \label{eq2}
{\it W_{xx}}+{\it W_{yy}}+{\frac {\partial }{\partial x}} \left(
2\,{ \it h_{x}}\,W \right) +{\frac {\partial }{\partial y}} \left(
2\,{\it h_{y}}\,W \right) =0
\end{equation}

by substitution \cite {Risken1989}

\begin{equation} \label{eq3}
Y \left( x,y \right) =W \left( x,y \right) {e^{h \left( x,y
\right) }}
\end{equation}

if $u$ and $h$ satisfy the condition
\begin{equation} \label{eq4}
{\it u}=-\Delta {\it h} +{{\it h_{x}}}^{2}+{{\it h_{y}}}^{2}
\end{equation}

The Fokker-Planck equation \eqref{eq2} has the conservation low
form that yields a pair of potential equations

\begin{equation} \label{eq5}
{\it W_{x}}+ 2\,{ \it h_{x}}\,W  -{\it Q_{y}} =0
\end{equation}
\begin{equation} \label{eq6}
{\it W_{y}}+ 2\,{ \it h_{y}}\,W  +{\it Q_{x}} =0
\end{equation}

In the paper \cite {Kudryavtsev2013} the special form of Darboux
transformation for the potential equations \eqref{eq5},
\eqref{eq6} was considered. As the potential variable is a
nonlocal variable for the Schr\"odinger equation that provides the
nonlocal Darboux transformation for the Schr\"odinger equation. It
was shown that this nonlocal transformation is a useful tool for
solvable two - dimensional stationary Schr\"odinger operators
obtaining. The consideration in the paper \cite {Kudryavtsev2013}
is restricted by the simple case $h=0$. In the present paper the
case of arbitrary $h$ is considered and relation of the nonlocal
Darboux transformation to the Moutard \cite {Moutard1878}
transformation is established.

\section{The nonlocal Darboux
transformation and its relation to the Moutard transformation}

Let us consider linear operator corresponding to the system of
equations \eqref{eq5}, \eqref{eq6}

\begin{equation*}
\hat L\left( h \left( x ,y \right)  \right) \, {\bf F}=
\begin{pmatrix} { 2\,{ \it h_{x}}+\frac {\partial }{\partial x} }  &
{ -\frac {\partial }{\partial y} } \\ {{2\,{ \it h_{y}}+\frac
{\partial }{\partial y}}}  & {\frac {\partial }{\partial x}}
\end{pmatrix}\,
\begin{pmatrix} F_1 \\ F_2 \end{pmatrix}
\end{equation*}

Consider Darboux transformation in the form

\begin{equation*}
\hat L_D \, {\bf F}=
\begin{pmatrix} { r_{11}-a_{11}\,\frac {\partial }{\partial x}-b_{11}\,\frac {\partial }{\partial y}  }  &
{  r_{12}-a_{12}\,\frac {\partial }{\partial x}-b_{12}\,\frac
{\partial }{\partial y} } \\ { r_{21}-a_{21}\,\frac {\partial
}{\partial x}-b_{21}\,\frac {\partial }{\partial y} }  & {
r_{22}-a_{22}\,\frac {\partial }{\partial x}-b_{22}\,\frac
{\partial }{\partial y} }
\end{pmatrix} \,
\begin{pmatrix} F_1 \\ F_2 \end{pmatrix}
\end{equation*}

If linear operators $\hat L$ and $\hat L_{D}$ hold the
intertwining relation

\begin{equation} \label{eq7}
\left( \hat L\left( h \left( x ,y \right) + s \left( x ,y
\right) \right)\hat L_{D} - \hat L_{D} \hat L\left( h \left( x ,y
\right) \right) \right) \, {\bf F}= 0
\end{equation}

for any $ {\bf F} \in  \mathcal{F} \supset Ker\left( \hat L\left(
h \right)\right)$ where  $Ker\left( \hat L\left( h
\right)\right)=\{{\bf F}:{\hat {L}}\left( h \right){\bf F}=0\}$,
then for any ${\bf F_s}\in Ker\left( \hat L\left( h
\right)\right)$
 the function $\tilde {\bf F} \left( x,y \right)=
\hat L_{D} {\bf F_s} \left( x,y \right)$ is a solution of the
equation ${\hat {L}}\left( {\tilde {h}} \right) \tilde {\bf F} =0$
\, with new potential $\tilde h = h+s$.

In the paper \cite {Kudryavtsev2013} it was proposed to consider
equation \eqref{eq7} on the following domain:
\begin{equation*}
{\mathcal{F}}_0=\{\,{\bf F}: {F_1}_{x}+ 2\,{ \it h_{x}}\, F_1  -
{F_2}_{y} =0\}
\end{equation*}

Taking into account this dependance of $F_1, F_2$ derivatives, the
equations for $s, r_{ij}, a_{ij}, b_{ij}$ can be obtained.

It should be noted that instead of restriction to solutions of
equation \eqref{eq5} domain ${\mathcal{F}}_0$ can be defined by
restriction to solutions of equation \eqref{eq6}. Because the
system of equations \eqref{eq5}, \eqref{eq6} has $x,y$ interchange
with change of $Q$ sign symmetry, one can simply use this symmetry
in resulting formulae.

The case $h=0$ was considered in the paper \cite {Kudryavtsev2013}
. Now let us consider the case of arbitrary $h$.

When solving equation \eqref{eq7} on the domain ${\mathcal{F}}_0$
the special situation arise in the case $s=-2h$ that corresponds
to $\tilde h = -h$. In this case we obtain from the equation
\eqref{eq7} the following Darboux transformation

\begin{equation} \label{eq9}
\hat L_D = {e^{2\,h \left( x,y \right) }}
\begin{pmatrix} { 0 }  &
{ 1 } \\ { - 1 }  & { 0 }
\end{pmatrix} \,
\end{equation}

By the formula \eqref{eq4} with $\tilde h = -h$ we obtain for the
new Schr\"odinger potential

\begin{equation} \label{eq10}
\tilde {u} \left( x ,y \right) = u\left( x ,y \right) + 2\,\Delta
\, h\left( x,y \right)
\end{equation}

Consider
\begin{equation} \label{eq11}
Y_h\left( x ,y \right)=e^{-h \left( x,y \right) }
\end{equation}

According to the formula \eqref{eq4}, $Y_h$ is a solution of the
Schr\"odinger equation with potential $u$.

Then we get another form of the formula \eqref{eq10}

\begin{equation} \label{eq12}
\tilde {u} \left( x ,y \right) = u\left( x ,y \right) - 2\,\Delta
\, \ln  \left( Y_h \left( x,y \right)  \right)
\end{equation}

This formula coincides with the formula of Moutard transformation
for the potential of the Schr\"odinger equation.

From the formula \eqref{eq9} and relation

\begin{equation} \label{eq13}
\tilde {Y} \left( x,y \right) =\tilde {W} \left( x,y \right) {e^{
\tilde {h} \left( x,y \right) }}
\end{equation}

we have for the new solutions of the Fokker-Planck and
Schr\"odinger equations

\begin{equation} \label{eq14}
\tilde {W} \left( x ,y \right) = {e^{2\,h \left( x,y \right) }}Q
\left( x,y \right)
\end{equation}

\begin{equation} \label{eq15}
\tilde {Y} \left( x ,y \right) = {e^{h \left( x,y \right) }}Q
\left( x,y \right)
\end{equation}

One can express $W, Q, h$ by equations \eqref{eq3}, \eqref{eq14},
\eqref{eq11} trough $Y, \tilde {Y}, Y_h$ and substitute to the
system of equations \eqref{eq5}, \eqref{eq6}. The result is

\begin{equation} \label{eq16}
{\frac {\partial }{\partial x}} \left( {\it Y_h} \left( x,y
\right) {\tilde {Y}} \left( x,y \right)  \right) + \left( {\it
Y_h} \left( x,y \right)  \right) ^{2}{\frac {\partial }{\partial
y}} \left( {\frac {{ \it Y} \left( x,y \right) }{{\it Y_h} \left(
x,y \right) }} \right)=0
\end{equation}

\begin{equation} \label{eq17}
{\frac {\partial }{\partial y}} \left( {\it Y_h} \left( x,y
\right) {\tilde {Y}} \left( x,y \right)  \right) - \left( {\it
Y_h} \left( x,y \right)  \right) ^{2}{\frac {\partial }{\partial
x}} \left( {\frac {{ \it Y} \left( x,y \right) }{{\it Y_h} \left(
x,y \right) }} \right)=0
\end{equation}

These formulae coincide with the formulae of the Moutard
transformation for the solution of the Schr\"odinger equation.

Thus the case $\tilde h = -h$ of the nonlocal Darboux
transformation provides the Moutard transformation.

The twofold application of the Moutard transformation can be
effective for obtaining nonsingular solvable potentials for the
Schr\"odinger equation \cite {Tsarev2008}. Now we give some
formulae of the twofold Moutard transformation in the form
convenient for later use.

Let us consider a solvable Schr\"odinger potential $u$ and select
two solutions $Y_{1}, Y_{2}$ of the Schr\"odinger equation with
this potential. Then according to \eqref{eq11} we choose
$Y_{h1}=Y_{1}$ and perform the Moutard transformation by the
change of $h_1$ sign. From \eqref{eq15} we have for the result of
$Y_{2}$ transformation ${\tilde {Y}}_{2}=Q_{12}/Y_{h1} $, where
$Q_{12}$, according to \eqref{eq5}, \eqref{eq6}, \eqref{eq3} and
\eqref{eq11}, satisfies the following system of equations:
\begin{equation} \label{eq18}
 {\frac {\partial Y_{2}}{\partial x}}  \, Y
_{1}  -   Y_{2} \, {\frac {\partial Y_{1}}{\partial x}}
 -{\frac {\partial Q_{12}}{\partial y}} =0, \,\,\,\,
{\frac {\partial Y_{2}}{\partial y}} \, Y_{1} -Y_{2}  {\frac
{\partial Y_{1}}{\partial y}} +{\frac {\partial Q_{12}}{\partial
x} } =0
\end{equation}

Then we choose $Y_{h2}={\tilde {Y}}_{2}$ and perform the Moutard
transformation by the change of $h_2$ sign. From \eqref{eq4} for
$h=-h_2$ taking into account \eqref{eq18} we get

\begin{multline} \label{eq19}
\tilde {\tilde {u}}=u + 4 \, {\left( {Q_{12}} \right) ^{-1}}
\left( {\frac {\partial {Y_{2}}}{\partial y}}  {\frac {\partial
{Y_{1}}}{\partial x}} - {\frac {\partial {Y_{2}}}{\partial x}}
{\frac {\partial {Y_{1}}}{\partial y}} \right)
\\
+ 2 \, {\left( {Q_{12}} \right) ^{-2}}\left( \left( {\frac
{\partial {Y_{2}}}{\partial x}}  {Y_{1}} -{Y_{2}} {\frac {\partial
{Y_{1}}}{\partial x}} \right) ^{2}+ \left(  {\frac {\partial
{Y_{2}}}{\partial y}} {Y_{1}} -{Y_{2}} {\frac {\partial
{Y_{1}}}{\partial y}} \right) ^{2} \right)
\end{multline}

For nonsingular $u, Y_{1}, Y_{2}$ one can see from this formula
that singularity of $\tilde {\tilde {u}}$ can arise from $Q_{12}$
zeros. Note that $Q_{12}$ is defined by \eqref{eq18} to within
arbitrary constant. In some cases, this constant allows to make
$Q_{12}$ of constant signs.

The system of equations \eqref{eq16}, \eqref{eq17} for $Y=Y_h$ has
a simple solution $\tilde {Y}=1/Y_h$. Therefore $\tilde {\tilde
{Y}}=1/Y_{h2}=Y_{1}/Q_{12} $ is an example of solution for the
Schr\"odinger equation with potential $\tilde {\tilde {u}}$.

Now let us return to the consideration of the equation
\eqref{eq7}. In the case $s$ not equal to $-2h$ from the equation
\eqref{eq7} on the domain ${\mathcal{F}}_0$ we obtain the
following Darboux transformation

\begin{equation} \label{eq20}
\hat L_D = {e^{- s \left( x,y \right) }}
\begin{pmatrix} { R_{1}-\frac {\partial }{\partial y}  }  &
{  R_{2} } \\ { -{s}_{x}-R_{2} }  & { -{s}_{y}+R_{1} -\frac
{\partial }{\partial y}}
\end{pmatrix} \,
\end{equation}

where $R_{1} =F_{1}/F, \, R_{2} =F_{2}/F$
\begin{multline} \label{eq21}
F =  2s_{{x}} \left( s_{{x}}+2h _{{x}} \right) +2s_{{y}} \left(
s_{{y}}+2h_{{y}} \right)
\\
\shoveleft{
F_{1} =
 \left( s_{{x}}+2h_{{x}} \right)  \left( -2 \left(
s_{{{\it xy}}}+h_{{{\it xy}}} \right) + \left( s_{{y}}-2h_{{y}}
\right) s_{{x}} \right) }
\\
\shoveright{+ \left(s_{{y}} +2h_{{y}} \right)  \left( s_{{{\it
xx}}}-s_{{{\it yy} }}-2h_{{{\it yy}}}+ \left( s_{{y}}-2h_{{y}}
\right) s_{{y}}
 \right)
}
\\
\shoveleft{
F_{2} =
 s_{{x}}s_{{{\it xx}}}+2\,s_{{y}} \left( s_{{{\it
xy}}}+h_{{{ \it xy}}} \right) -s_{{x}} \left( s_{{{\it
yy}}}+2h_{{{\it yy}}}
 \right)
 }
\\
 - \left( s_{{x}}+2h_{{x}} \right){s_{{x}}}^{2} - \left( s_{{y}} +2
h_{{y}}\right) s_{{x}}s_{{y}}
\end{multline}

and $s$ satisfies the following system of two nonlinear differential
equations:
\begin{multline} \label{eq22}
 \left( {s_{{x}}}^{3}+2\,{s_{{x}}}^{2}h_{{x}}+ \left( 2\,s_{{y}}h_{{y}
}+{s_{{y}}}^{2} \right) s_{{x}} \right) s_{{{\it xxx}}}
\\
\shoveleft{+ \left(
 \left( s_{{y}}-2\,h_{{y}} \right) {s_{{x}}}^{2}+ \left( -4\,h_{{y}}h_
{{x}}+2\,h_{{x}}s_{{y}} \right)
s_{{x}}+{s_{{y}}}^{3}-4\,{h_{{y}}}^{2} s_{{y}} \right) s_{{{\it
xxy}}} }
\\
\shoveleft{+ \left( {s_{{x}}}^{3}+6\,{s_{{x}}}^{2} h_{{x}}+ \left(
8\,{h_{{x}}}^{2}+2\,s_{{y}}h_{{y}}+{s_{{y}}}^{2}
 \right) s_{{x}}+8\,h_{{x}}s_{{y}}h_{{y}}+4\,h_{{x}}{s_{{y}}}^{2}
 \right) s_{{{\it xyy}}}  }
\\
\shoveleft{ + \left(  \left( 2\,h_{{y}}+s_{{y}} \right) {s
_{{x}}}^{2}+ \left( 4\,h_{{y}}h_{{x}}+2\,h_{{x}}s_{{y}} \right)
s_{{x}
}+4\,h_{{y}}{s_{{y}}}^{2}+{s_{{y}}}^{3}+4\,{h_{{y}}}^{2}s_{{y}}
 \right) s_{{{\it yyy}}}  }
\\
\shoveleft{ + \left( 2s_{{y}}h_{{y}}+{s_{{y}}}^{2}-{s_{{ x}}}^{2}
\right) {s_{{{\it xx}}}}^{2}  +  2\left(  \left( -2s_{{
y}}+h_{{y}} \right) s_{{x}}+2h_{{y}}h_{{x}}-h_{{x}}s_{{y}}
 \right)s_{{{\it xx}}} s_{{{\it xy}}}  }
\\
\shoveleft{ +  2\left(
-h_{{x}}s_{{x}}+2{h_{{y}}}^{2}+s_{{y}}h_{{y}} \right) s_{{{\it
xx}}} s_{{{\it yy}}} + 4\left( s_{{y}}h_{{y}}-h_{{x}}s_{{x}
}-2{h_{{x}}}^{2} \right) {s_{{{\it xy}}}}^{2}  }
\\
\shoveleft{+2\left(-{h_{{{\it xx}}}s_{{x}}}^{2} + 2\left(
-h_{{{\it yy}}}h_{{x}}-s_{{y}}h_{{{\it xy}}}+h_{{{ \it
xy}}}h_{{y}} \right) s_{{x}}\right) s_{{{\it xx}}}  }
\\
\shoveleft{-2\left(h_{{{\it yy}}}{s_{{y}}}^{2}+2
 \left( h_{{{\it yy}}}h_{{y}}+h_{{{\it xy}}}h_{{x}} \right) s_{
{y}} \right) s_{{{\it xx}}}  }
\\
\shoveleft{ + \left(
 \left( -6\,h_{{y}}-4\,s_{{y}} \right) s_{{x}}-10\,h_{{x}}s_{{y}}-12\,
h_{{y}}h_{{x}} \right) s_{{{\it xy}}} s_{{{\it yy}}}  }
\\
\shoveleft{ + \left(  \left( -4\,h_{{{\it xx} }}-4\,h_{{{\it yy}}}
\right) s_{{y}}-4\,h_{{{\it yy}}}h_{{y}}-4\,h_{{{ \it xy}}}h_{{x}}
\right) s_{{x}} s_{{{\it xy}}}  }
\\
\shoveleft{+ \left( \left( -12\,h_{{{\it yy}}}h_{{x}}+12
\,h_{{{\it xy}}}h_{{y}} \right) s_{{y}}-8\,h_{{x}}h_{{{\it
yy}}}h_{{y} }-8\,{h_{{x}}}^{2}h_{{{\it xy}}} \right) s_{{{\it
xy}}}  }
\\
\shoveleft{+ \left( -4\,{h
_{{y}}}^{2}+{s_{{x}}}^{2}+2\,h_{{x}}s_{{x}}-4\,s_{{y}}h_{{y}}-{s_{{y}}
}^{2} \right) {s_{{{\it yy}}}}^{2}  }
\\
\shoveleft{+ \left(  \left( 4\,h_{{{\it yy}}}+2 \,h_{{{\it xx}}}
\right) {s_{{x}}}^{2}+ \left( -8\,h_{{{\it xy}}}h_{{y
}}+8\,h_{{{\it yy}}}h_{{x}}-4\,s_{{y}}h_{{{\it xy}}} \right)
s_{{x}} \right) s_{{{\it yy}}}  }
\\
\shoveleft{+\left( -2 \,h_{{{\it yy}}}{s_{{y}}}^{2}+ \left(
-8\,h_{{{\it xy}}}h_{{x}}-8\,h_{ {{\it yy}}}h_{{y}} \right)
s_{{y}}-8\,h_{{x}}h_{{{\it xy}}}h_{{y}}-8\, h_{{{\it
yy}}}{h_{{y}}}^{2} \right) s_{{{\it yy}}}  }
\\
\shoveleft{-{s_{{x}}}^{6}-6\,h_ {{x}}{s_{{x}}}^{5}-3\left(
4{h_{{x}}}^{2}+2s_{{y}}h_{{y}}+{s_ {{y}}}^{2} \right)
{s_{{x}}}^{4}-4 \left( 2{h_{{x}}}^{3}+6h_{{x}}
s_{{y}}h_{{y}}+3h_{{x}}{s_{{y}}}^{2} \right) {s_{{x}}}^{3}  }
\\
\shoveleft{- \left( 3{s_{{y}}}^{4}+12h_{{y}}{s_{{y}}}^{3}+12\left(
{h_{{y}}}^{2}+{h_{{x}}}^{2} \right) {s_{{y}}}^{2}- 2\left(
-12{h_{{x}}}^{2}h_{{ y}}+h_{{{\it yyy}}}+h_{{{\it xxy}}} \right)
s_{{y}}\right) {s_{{x}}}^{2}  }
\\
\shoveleft{+4\left(h_{{{\it yyy}}}h_{{y}}+h_{{{\it yy}}}h_{{{\it
xx}}}+{h_{{{\it yy}}}}^{2}+ h_{{{\it xyy}}}h_{{x}} \right)
{s_{{x}}}^{2}  }
\\
\shoveleft{-6\left( h_{{x}}{s_
{{y}}}^{4}+4h_{{x}}h_{{y}}{s_{{y}}}^{3}+4h_{{x}}{h_{{y}}}^{2}{s_
{{y}}}^{2}\right)s_{{x}}  }
\\
\shoveleft{-4\left( h_{{{\it yy}}}h_{{{\it xy}}}-h_{{x}}h_{{{ \it
yyy}}}+h_{{{\it xy}}}h_{{{\it xx}}}-h_{{{\it xxy}}}h_{{x}}
 \right) s_{{y}}s_{{x}}  }
\\
\shoveleft{+\left(8\,h_{{x}}h_{{{\it
yyy}}}h_{{y}}+8\,{h_{{x}}}^{2}h_{{ {\it xyy}}}-8\,h_{{{\it
xy}}}h_{{{\it yy}}}h_{{y}}+8\,h_{{x}}{h_{{{ \it yy}}}}^{2} \right)
s_{{x}}-{s_{{y}}}^{6}-6\,h_{{y}}{s_{{y}}}^{5}  }
\\
\shoveleft{- 12\,{h_{{y}}}^{2}{s_{{y}}}^{4}  + 2\left( h_{{{\it
yyy}}}-4{h_{{y}}} ^{3}+h_{{{\it xxy}}} \right) {s_{{y}}}^{3}+
4\left( h_{{{\it xxy} }}h_{{y}}+2h_{{{\it yyy}}}h_{{y}}+h_{{{\it
xyy}}}h_{{x}} \right) {s_{{y}}}^{2}  }
\\
+ \left( -8\,h_{{x}}h_{{{\it xy}}}h_{{{\it yy}}}+8\,h_{{{ \it
yyy}}}{h_{{y}}}^{2}+8\,h_{{x}}h_{{{\it xyy}}}h_{{y}}+8\,{h_{{{\it
xy}}}}^{2}h_{{y}} \right) s_{{y}}=0
\end{multline}

\begin{multline} \label{eq23}
 \left(  \left( 2\,h_{{y}}+s_{{y}} \right) {s_{{x}}}^{2}+ \left( 4\,h_
{{y}}h_{{x}}+2\,h_{{x}}s_{{y}} \right)
s_{{x}}+4\,h_{{y}}{s_{{y}}}^{2}
+{s_{{y}}}^{3}+4\,{h_{{y}}}^{2}s_{{y}} \right) s_{{{\it xxx}}}
\\
\shoveleft{+
 \left( -{s_{{x}}}^{3}-6\,{s_{{x}}}^{2}h_{{x}}+ \left( -2\,s_{{y}}h_{{
y}}-8\,{h_{{x}}}^{2}-{s_{{y}}}^{2} \right)
s_{{x}}-4\,h_{{x}}{s_{{y}}} ^{2}-8\,h_{{x}}s_{{y}}h_{{y}} \right)
s_{{{\it xxy}}}  }
\\
\shoveleft{+ \left( \left( s_{{y}}-2\,h_{{y}} \right)
{s_{{x}}}^{2}+ \left( -4\,h_{{y}}h_{{x}}+2 \,h_{{x}}s_{{y}}
\right) s_{{x}}+{s_{{y}}}^{3}-4\,{h_{{y}}}^{2}s_{{y}}
 \right) s_{{{\it xyy}}}   }
\\
\shoveleft{ + \left( -{s_{{x}}}^{3}-2\,{s_{{x}}}^{2}h_{{x} }+
\left( -{s_{{y}}}^{2}-2\,s_{{y}}h_{{y}} \right) s_{{x}} \right)
s_{ {{\it yyy}}}   }
\\
\shoveleft{+ \left(  \left( -2\,s_{{y}}-4\,h_{{y}} \right)
s_{{x}}-4 \,h_{{y}}h_{{x}}-2\,h_{{x}}s_{{y}} \right) {s_{{{\it
xx}}}}^{2}   }
\\
\shoveleft{+ \left(
10\,h_{{x}}s_{{x}}-6\,s_{{y}}h_{{y}}+2\,{s_{{x}}}^{2}+
8\,{h_{{x}}}^{2}-4\,{h_{{y}}}^{2}-2\,{s_{{y}}}^{2} \right)
s_{{{\it xy }}}s_{{{\it xx}}}   }
\\
\shoveleft{+ \left(
2\,s_{{x}}h_{{y}}+2\,h_{{x}}s_{{y}}+4\,h_{{y}}h_{{x}}
 \right) s_{{{\it yy}}} s_{{{\it xx}}}   }
\\
\shoveleft{ +\left(2\,h_{{{\it xy}}}{s_{{x}}}^{2}+ \left(
 \left( -2\,h_{{{\it xx}}}+2\,h_{{{\it yy}}} \right) s_{{y}}+12\,h_{{{
\it xy}}}h_{{x}}-4\,h_{{{\it xx}}}h_{{y}}+8\,h_{{{\it yy}}}h_{{y}}
 \right) s_{{x}}\right)s_{{{\it xx}}}   }
\\
\shoveleft{ +\left(-2\,h_{{{\it xy}}}{s_{{y}}}^{2}+ \left(
4\,h_{{{\it yy }}}h_{{x}}-4\,h_{{{\it xy}}}h_{{y}} \right)
s_{{y}}+8\,h_{{x}}h_{{{ \it yy}}}h_{{y}}+8\,{h_{{x}}}^{2}h_{{{\it
xy}}} \right) s_{{{\it xx}}}    }
\\
\shoveleft{ + 4\left(
s_{{x}}h_{{y}}+2h_{{y}}h_{{x}}+h_{{x}}s_{{y}}
 \right) {s_{{{\it xy}}}}^{2}+  2\left( h_{{x}}s_{{x}}-{s_
{{y}}}^{2}+{s_{{x}}}^{2}+2{h_{{y}}}^{2}+s_{{y}}h_{{y}}
 \right) s_{{{\it yy}}}s_{{{\it xy
}}}     }
\\
\shoveleft{ + 4\left(h_{{{\it xy}}}h_{{y}}  -h_{{{\it
yy}}}h_{{x}}\right) s_{{x}}s_{{{\it xy }}}-4\left( h_{{{\it
xx}}}+h_{{{\it yy }}} \right) {s_{{y}}}^{2}s_{{{\it xy }}}   }
\\
\shoveleft{+ 4\left( 3h_{{{\it xy}}}h_{{x}}-2h_{{{ \it
xx}}}h_{{y}}+h_{{{\it yy}}}h_{{y}} \right) s_{{y}}s_{{{\it xy }}}
}
\\
\shoveleft{+\left( 8\,h_{{x}}h_ {{{\it xy}}}h_{{y}}+8\,h_{{{\it
yy}}}{h_{{y}}}^{2} \right) s_{{{\it xy }}}+ \left(
2\,h_{{y}}+2\,s_{{y}} \right) s_{{x}}{s_{{{\it yy}}}}^{2}    }
\\
\shoveleft{+
 \left( 2\,h_{{{\it xy}}}{s_{{x}}}^{2}+ \left(  \left( 6\,h_{{{\it yy}
}}+2\,h_{{{\it xx}}} \right) s_{{y}}+4\,h_{{{\it
yy}}}h_{{y}}+4\,h_{{{ \it xx}}}h_{{y}} \right) s_{{x}}-2\,h_{{{\it
xy}}}{s_{{y}}}^{2}
 \right) s_{{{\it yy}}}     }
\\
\shoveleft{ + \left( -2\,h_{{{\it yyy}}}-2\,h_{{{\it xxy}}}
 \right) {s_{{x}}}^{3}+ \left( -4\,h_{{x}}h_{{{\it yyy}}}-8\,h_{{{\it
xxy}}}h_{{x}}-4\,h_{{y}}h_{{{\it xyy}}} \right) {s_{{x}}}^{2}
}
\\
\shoveleft{+ \left(
 \left( -2\,h_{{{\it yyy}}}-2\,h_{{{\it xxy}}} \right) {s_{{y}}}^{2}+
 \left( -4\,h_{{{\it xxy}}}h_{{y}}-4\,h_{{{\it yyy}}}h_{{y}}+4\,{h_{{{
\it yy}}}}^{2}+4\,h_{{{\it yy}}}h_{{{\it xx}}} \right)
s_{{y}}\right) s_{{x}}     }
\\
\shoveleft{+\left(-8\,h_{{ x}}h_{{{\it xy}}}h_{{{\it
yy}}}-8\,h_{{x}}h_{{{\it xyy}}}h_{{y}}-8\,{h _{{x}}}^{2}h_{{{\it
xxy}}}+8\,h_{{{\it yy}}}h_{{y}}h_{{{\it xx}}}
 \right) s_{{x}}     }
\\
\shoveleft{ + \left( -4\,h_{{y}}h_{{{\it xyy}}}-4\,h_{{{\it yy}}}h
_{{{\it xy}}}-4\,h_{{{\it xxy}}}h_{{x}}-4\,h_{{{\it xy}}}h_{{{\it
xx}} } \right) {s_{{y}}}^{2}     }
\\
+ \left( -8\,{h_{{y}}}^{2}h_{{{\it xyy}}}-8\,h_
{{y}}h_{{x}}h_{{{\it xxy}}}+8\,h_{{x}}{h_{{{\it
xy}}}}^{2}-8\,h_{{{ \it xx}}}h_{{y}}h_{{{\it xy}}} \right) s_{{y}}
=0
\end{multline}

By the formula \eqref{eq4} with $\tilde h = h+s$ we obtain for the
new Schr\"odinger potential

\begin{equation} \label{eq24}
\tilde {u} = u  -\Delta s+2\,h_{{x}}s_{{x}}+{s_{{x}}}^{2}+2\,s_{{
y}}h_{{y}}+{s_{{y}}}^{2}
\end{equation}

From the formula \eqref{eq20} we have for the new solution of the
Fokker-Planck equation

\begin{equation} \label{eq25}
\tilde {W}  = e^{- s}\left( R_{1}\,W- \frac {\partial W}{\partial
y}  + R_{2}\,Q \right)
\end{equation}

From the formulae \eqref{eq25}, \eqref{eq13} and \eqref{eq3} we
have for the new solution of the Schr\"odinger equation

\begin{equation} \label{eq26}
\tilde {Y}  =  \left( R_{{1}}  +h_{{y}} \right) Y  -\frac
{\partial Y}{\partial y}   +{e^{h}}R_{{2}}\, Q
\end{equation}

Where $Q$, according to \eqref{eq5}, \eqref{eq6}, \eqref{eq3} and
\eqref{eq11}, satisfies the following system of equations
\begin{equation} \label{eq27}
 {\frac {\partial Y}{\partial x}}  \, Y
_{{h}}  -   Y \, {\frac {\partial Y_{{ h}}}{\partial x}}
 -{\frac {\partial Q}{\partial y}} =0
\end{equation}

\begin{equation} \label{eq28}
  {\frac {\partial Y}{\partial y}} \, Y
_{{h}} -Y  {\frac {\partial Y_{{h}}}{\partial y}} +{\frac
{\partial Q}{\partial x} } =0
\end{equation}

The Darboux transformation \eqref{eq26} is a nonlocal
transformation as the potential variable $Q$ is a nonlocal
variable connected with $Y$ by the system \eqref{eq27},
\eqref{eq28}.

\section{Application of
the nonlocal Darboux transformation }

In the simple case $h=0$ the system of equations \eqref{eq22},
\eqref{eq23} for $B=\exp \left( -s \right)$ has the form

\begin{multline} \label{eq29}
 - \left( 2\,B{\it
B_y}\,{\it B_{xy}}+B{\it B_x}\, \left( {\it B_{xx}}-{\it B_{yy}}
 \right) +{\it B_x} \left( {{\it B_x}}^{2}+{{\it B_y}}^{2}
 \right)  \right)  \left( {\it B_{xx}}+{\it B_{yy}} \right)
\\
\shoveright{ +B \left( {{\it B_x}}^{2}+{{\it B_y}}^{2} \right)
{\frac {\partial }{
\partial x}} \left( {\it B_{xx}}+{\it B_{yy}} \right) =0}
\\
\shoveleft{ - \left( 2\,B{\it B_x}\,{\it B_{xy}}-B{\it B_y}\,
\left( {\it B_{xx}}-{\it B_{yy}}
 \right) +{\it B_y}\, \left( {{\it B_x}}^{2}+{{\it B_y}}^{2}
 \right)  \right)  \left( {\it B_{xx}}+{\it B_{yy}} \right)}
 \\
 +B \left( {{\it B_x}}^{2}+{{\it B_y}}^{2} \right) {\frac
{\partial }{
\partial y}} \left( {\it B_{xx}}+{\it B_{yy}} \right)
 =0
\end{multline}

These equations were considered in the paper \cite
{Kudryavtsev2013}. If $B$ is a solution of equations \eqref{eq29}
then $1/B$ and $CB$, where $C$ is an arbitrary constant, are
solutions as well. It is obvious from the form of equations
\eqref{eq29} that any solution of the Laplace equation $\Delta B =
0$ is the solution of these equations.

The system of equations \eqref{eq29} can be integrated. In the
case $\Delta B$ is not zero we obtain

\begin{equation} \label{eq30}
 \left(  {B} ^{2}-K \right)  \left( {\it B_{xx}}+{\it B_{yy}} \right) -2\,B
  \left(  {{\it B_x}}^{2}+{{\it B_y}}^{2} \right)=0
\end{equation}

or in the other form

\begin{equation} \label{eq31}
{\frac { {B} ^{4} \Delta  \left(  1/B \right)  }
{\Delta B}}=-K
\end{equation}

where $K$ is an arbitrary constant. It is obvious that if $B$ is a
solution of equation \eqref{eq31} with constant $K_B$ then $1/B$
is a solution of this equation with constant $1/K_B$.

According to equation \eqref{eq4}, the initial potential $u$ of the Schr\"odinger equation
corresponding to $h=0$ is $u=0$.
The new potential of Schr\"odinger equation corresponds to $s=-\ln
\left( B \right)$ and  is given by

\begin{equation} \label{eq32}
\tilde {u}  =  \Delta B/B
\end{equation}

Note that according to formula  \eqref{eq32} $B$ is an example of
solution for the Schr\"odinger equation with new potential $\tilde
{u}$.

The solution $B_L$ of the Laplace equation provide $\tilde {u}=0$
\eqref{eq32}. Taking $B=1/B_L$ one
obtains nontrivial $\tilde {u}$, but potentials of this kind have
singularities:

\begin{equation*}
\tilde {u} =2\, \left( {\it B_L}  \right) ^{-2}\left( \left( {
\frac {\partial {\it B_L} }{\partial x}} \right) ^ {2}+\left(
{\frac {\partial {\it B_L} }{\partial y}} \right) ^{2}\right)
\end{equation*}

Note that this formula coincides with the formula \eqref{eq12} for
the Moutard transformation of the potential where $u=0, Y_h=B_L$.

Let us consider the following ansatz for $B$

\begin{equation*}
B \left( x,y \right) =F \left( {x}^{2}+{y}^{2} \right)
\end{equation*}

This ansatz provides the following solution of equation
\eqref{eq31}

\begin{equation} \label{eq33}
B_r \left( x,y \right) =-{\frac { \sqrt {K}\left(  \left(
{x}^{2}+{y}^{2} \right) ^{{\it C_1}}-{\it C_2}
 \right) }{ \left( {x}^{2}+{y}^{2} \right) ^{{\it C_1}}+{\it
C_2}}}
\end{equation}

where ${\it C_1},{\it C_2}$ are arbitrary constants.

The solution $B_r$ provides by the formula \eqref{eq32}

\begin{equation} \label{eq34}
{\tilde {u}} =-8\,{\frac {{\it C_2}\,{{\it C_1}}^{2} \left(
{x}^{2}+{y}^{2} \right) ^{ {\it C_1}-1}}{ \left(  \left(
{x}^{2}+{y}^{2} \right) ^{{\it C1}}+{\it C_2} \right) ^{2}}}
\end{equation}

This solvable potential is nonsingular if ${\it C_1} \ge 1, {\it
C_2}>0$. For ${\it C_1}=1, 2, 3, 4$ this potential is the special
case of potentials considered in \cite {Kudryavtsev2013}.

Consider for example ${\it C_1} =3/2, {\it C_2}=1$. In this case
we have

\begin{equation} \label{eq35}
B ={\frac { \left( {x}^{2}+{y}^{2} \right) ^{3/2}-1  }{ \left(
{x}^{2}+{y}^{2} \right) ^{3/2}+1}}
\end{equation}

Consider two solutions of the Laplace equation
\begin{equation*}
Y_{L1} ={\frac {x}{{x}^{2}+{y}^{2}}}, \,\,\, Y_{L2} ={\frac
{y}{{x}^{2}+{y}^{2}}}
\end{equation*}

From \eqref{eq27}, \eqref{eq28} with $Y_h=1$ we have for
$Y=Y_{L1}$ and $Y=Y_{L2}$
\begin{equation*}
Q_{L1} =-{\frac {y}{{x}^{2}+{y}^{2}}}+{\it C_{L1}}, \,\,\, Q_{L2}
={\frac {x}{{x}^{2}+{y}^{2}}}+{\it C_{L2}}
\end{equation*}

where ${\it C_{L1}},{\it C_{L2}}$ are arbitrary constants.

From \eqref{eq26} with $h=0$ and $B$ from \eqref{eq35} we have for
$Y=Y_{L1}, Q=Q_{L1}$ and $Y=Y_{L2}, Q=Q_{L2}$

\begin{equation} \label{eq36}
{\tilde {Y}}_{L1} = {\frac {-2\,xy \left( 7\, \left(
{x}^{2}+{y}^{2} \right) ^{3/2}+1 \right) }{ \left( {x}^{2}+{y}^{2}
\right) ^{2} \left(  \left( {x}^{2}+{y}^{2} \right) ^{3/2}+1
\right) }}+{\frac {{\it C_{L1}} \, x \left( 5\, \left(
{x}^{2}+{y}^{2} \right) ^{3/2}-1 \right) }{ \left( {x}^{2}+{y}
^{2} \right) \left( \left( {x}^{2}+{y}^{2} \right) ^{3/2}+1
\right) }}
\end{equation}

\begin{equation} \label{eq37}
{\tilde {Y}}_{L2} = {\frac { \left( {x}^{2}-{y}^{2} \right) \left(
7\, \left( {x}^{ 2}+{y}^{2} \right) ^{3/2}+1 \right) }{ \left(
{x}^{2}+{y}^{2} \right) ^{2} \left(  \left( {x}^{2}+{y}^{2}
\right) ^{3/2}+1 \right) }}+{ \frac {{\it C_{L2}} \, x \left( 5\,
\left( {x}^{2}+{y}^{2} \right) ^{3/2}- 1 \right) }{ \left(
{x}^{2}+{y}^{2} \right)  \left(  \left( {x}^{2}+{y }^{2} \right)
^{3/2}+1 \right) }}
\end{equation}

These functions are examples of solutions for the Schr\"odinger
equation with solvable potential
\begin{equation} \label{eq38}
{\tilde {u}} =-18\,{\frac {\sqrt {{x}^{2}+{y}^{2}}}{ \left( \left(
{x}^{2}+{y}^{2} \right) ^{3/2}+1 \right) ^{2}}}
\end{equation}

Now the Moutard transformation can be applied to the Schr\"odinger
equation with potential \eqref{eq38}. A single Moutard
transformation can provide singular solvable potential. To obtain
new nonsingular solvable potential for the Schr\"odinger equation
from the solvable potential \eqref{eq38}, let us apply twofold
Moutard transformation . Let us consider $u$ from \eqref{eq38} and
select two solutions $Y_{1}, Y_{2}$ of the Schr\"odinger equation
equal to functions \eqref{eq36}, \eqref{eq37} with
$C_{L1}=C_{L2}=0$. From the equation \eqref{eq18} we obtain

\begin{equation} \label{eq39}
Q_{12}=
 {\frac {49\, \left( {x}^{2}+{y}^{2} \right) ^{3/2}+1}{ 2 \, \left( {x}
^{2}+{y}^{2} \right) ^{2} \left(  \left( {x}^{2}+{y}^{2} \right)
^{3/2 }+1 \right) }}+C/2
\end{equation}
where $C$ is an arbitrary constant. From the equation \eqref{eq19}
we obtain

\begin{equation} \label{eq40}
\tilde {\tilde {u}}= {\frac {-2\, r \left(
441+9\,{C}^{2}{r}^{8}+2\,Cr \left( 392\,{r}^{6}+ 49\,{r}^{3}+8
\right)  \right) }{ \left( C{r}^{4} \left( {r}^{3}+1
 \right) +49\,{r}^{3}+1 \right) ^{2}}}
\end{equation}
where $r=\sqrt {{x}^{2}+{y}^{2}}$.

This solvable potential is nonsingular if $C \ge 0$. The example
of solution for the Schr\"odinger equation with potential
\eqref{eq40} is
\begin{equation} \label{eq41}
\tilde {\tilde {Y}}=Y_{1}/Q_{12}= {\frac { -4\,xy\left( 7\, \left(
{x}^{2}+{y}^{2} \right) ^{3/2}+1 \right) }{ C \left(
{x}^{2}+{y}^{2} \right) ^{2} \left(  \left( {x}^{ 2}+{y}^{2}
\right) ^{3/2}+1 \right) +49\, \left( {x}^{2}+{y}^{2}
 \right) ^{3/2}+1}}
\end{equation}

The solutions of equations \eqref{eq22}, \eqref{eq23} can be
obtained not only for $h=0$. For example let us consider
$h(x,y)=H(y)$ and use the ansatz $s(x,y)=-2\,H(y)+S(x)$. That
reduces the equations \eqref{eq22}, \eqref{eq23} to the ordinary
differential equation
\begin{equation} \label{eq42}
S_{{x}}S_{{{\it xxx}}}-{S_{{{\it xx}}}}^{2}-{S_{{x}}}^{4}=0
\end{equation}

The solution of the equation \eqref{eq42} is
\begin{equation} \label{eq43}
S(x)=\ln  \left( {\frac {\exp \left( {\it C_1}\,x \right)-{\it
C_2}}{\exp \left( {\it C_1}\,x \right) +{ \it C_2}}} \right) +{\it
C_3}
\end{equation}
where ${\it C_1},{\it C_2},{\it C_3}$ are arbitrary constants.

Thus from any solvable potential
\begin{equation} \label{eq44}
u_H=-{\frac {d^{2} H \left( y \right)}{d{y}^{2}}} + \left( {\frac
{d H \left( y \right)}{dy}}  \right) ^{2}
\end{equation}
we obtain the new solvable potential
\begin{equation} \label{eq45}
{\tilde {u}}_{H}=u_H + 2\,{\frac {d^{2} H \left( y
\right)}{d{y}^{2}}} + {\frac {2\,{\it C_2}\,{{\it C_1}}^{2} \exp
\left({\it C_1}\,x\right)}{ \left( \exp \left({\it
C_1}\,x\right)-{\it C_2} \right) ^{2}}}
\end{equation}

Consider $H(y)=-\ln \left( \sin \left( y \right) \right)$ and
choose the solution \eqref{eq43} in the form $S(x)=\ln  \left(
\tanh \left( p \left( x-{\it x_0} \right)  \right) \right)$ where
$p,{\it x_0}$ are arbitrary constants. In this case the initial
solvable potential is $u_H=-1$ and
\begin{equation} \label{eq46}
{\tilde {u}}_{H}=-1+2\, \left( \sin \left( y \right)  \right)
^{-2}+2\,{p}^{2}\left( \sinh \left( p \left( x-{\it x_0} \right)
\right) \right) ^{-2}
\end{equation}
This potential has singularities. To obtain new nonsingular
solvable potential for the Schr\"odinger equation from the
solvable potential \eqref{eq46}, let us apply Moutard
transformation. For the solution $\sin \left( x \right)$ of the
Schr\"odinger equation with potential $u_H=-1$ by the formulae
\eqref{eq27}, \eqref{eq28} we obtain $Q= C -\cos \left( y \right)
\cos \left( x \right)$ where $C$ is an arbitrary constant. The
formula \eqref{eq26} provides the solution of the Schr\"odinger
equation with potential \eqref{eq46}
\begin{equation} \label{eq47}
{\tilde {Y}}_p ={\frac { \left( p \left( C - \cos \left( y \right)
\cos \left( x \right) \right) -\cos \left( y \right) \sin \left( x
\right) \tanh \left( p \left( x-{\it x_0} \right) \right) \right)
}{\sin \left( y \right) \tanh \left( p \left( x-{\it x_0} \right)
\right) }}
\end{equation}
We choose $Y_h={\tilde {Y}}_p$ and perform the Moutard
transformation for the potential \eqref{eq46}. The new solvable
potential is
\begin{equation} \label{eq48}
-1+ {\frac { 2 \, \left( {\it f_1}   \left( \cosh \left( p \left(
x-{\it x_0} \right)  \right)  \right) ^{-2}+{\it f_2} +{\it f_3}
\tanh \left( p \left( x-{\it x_0} \right)  \right) \right) }{
\left( p \left(C -\cos \left( y \right) \cos \left( x \right)
\right) -\cos \left( y \right) \sin \left( x \right) \tanh \left(
p \left( x-{\it x_0} \right) \right) \right) ^{2}}}
\end{equation}
where
\begin{multline*}
{\it f_1} =  -{p}^{2} \left( {p}^{2}+1 \right)  \left(C -\cos
\left( y \right) \cos \left( x \right)  \right)
^{2}+{p}^{2}{C}^{2}-\left( {p} ^{2}+1 \right)  { \cos \left( y
\right)  } ^{2}-{ \sin \left( x \right) } ^{2}
\\
\shoveleft{ {\it f_2} = -2\,{p}^{2}C \cos \left( y \right) \cos
\left( x \right) +1+ \left( {p}^{2}-1 \right){ \cos \left( x
\right)  } ^{2}  + \left( {p}^{2}+1 \right)  { \cos \left( y
\right)  } ^{2}
 }
\\
\shoveleft{ {\it f_3} = 2\,p \, \sin \left( x \right)  \left( \cos
\left( x \right) -C\cos \left( y \right)  \right)
 }
\\
\end{multline*}
This potential is obviously nonsingular for $p>0, \, C>1/p+1$. The
function $1/{\tilde {Y}}_p$ is an example of solution for the
Schr\"odinger equation with potential \eqref{eq48}.

\section{Results and Discussion}

In the past decades substantial progress in the description of the
exact solutions for nonlinear partial differential equations and
linear partial differential equations with variable coefficients
has taken place. The useful tool for solving one - dimensional
Schr\"odinger equation is the Darboux transformation \cite
{Matveev1991}. Some generalizations of the Darboux transformation
for two - dimensional case were proposed, including operators of
second order in derivatives \cite {Andrianov1995}, \cite
{Ioffe2004}. Examples of solvable two � dimensional models were
obtained in the framework of these generalized Darboux
transformations (see for example \cite {Andrianov2012}, \cite
{Bardavelidze2013} and references therein).  Some examples of
exactly solvable two - dimensional stationary Schr\"odinger
operators with smooth rational potentials decaying at infinity
were obtained in the paper \cite {Tsarev2008} by application of
the Moutard transformation. In the past years progress was made in
the symmetry group analysis of differential equations by extending
the spaces of symmetries of a given partial differential equations
system to include nonlocal symmetries \cite {Ibragimov1991}, \cite
{Bluman2010}. In the present paper the inclusion of nonlocal
variables in the Darboux transformation for the 2D stationary
Schr\"odinger equation is considered.

In the paper \cite {Kudryavtsev2013} the Fokker-Planck equation
associated with the two - dimensional stationary Schr\"odinger
equation was considered. The Fokker-Planck equation has the
conservation low form that yields a pair of potential equations.
The special form of the Darboux transformation for these potential
equations was introduced. As the potential variable is a nonlocal
variable that provides the nonlocal Darboux transformation for the
Schr\"odinger equation. It was shown that this nonlocal
transformation is a useful tool for obtaining solvable two -
dimensional stationary Schr\"odinger operators. The consideration
in the paper \cite {Kudryavtsev2013} is restricted by the simple
case $h=0$. In the present paper the case of arbitrary $h$ is
considered and relation of the nonlocal Darboux transformation to
the Moutard transformation is established. It is shown that the
special case $\tilde h = -h$ of the nonlocal Darboux
transformation provides the Moutard transformation. New examples
of solvable two - dimensional stationary Schr\"odinger operators
with smooth potentials are obtained as an application of the
nonlocal Darboux transformation.

\end{document}